\documentclass[aps,prc,twocolumn,superscriptaddress,nofootinbib]{revtex4-1} 

\usepackage{epsfig}
\usepackage{color}
\usepackage[latin1]{inputenc}
\usepackage{float,amsmath}
\usepackage{graphicx}
\usepackage{comment}
\usepackage{ulem}

\begin{document}

\title{Initial state geometry and fluctuations\\ in Au+Au, Cu+Au and U+U collisions at RHIC}
\author{Bj\"orn Schenke}
\affiliation{Physics Department, Brookhaven National Laboratory, Upton, NY 11973, USA}
\author{Prithwish Tribedy}
\affiliation{Variable Energy Cyclotron Centre, 1/AF Bidhan Nagar, Kolkata 700064, India}
\author{Raju Venugopalan}
\affiliation{Physics Department, Brookhaven National Laboratory, Upton, NY 11973, USA}

\begin{abstract}
We study within the IP-Glasma and two-component MC-Glauber models the effects of initial state geometry and fluctuations on multiplicities and eccentricities for 
several collision species at the Relativistic Heavy Ion Collider (RHIC). These include copper-gold (Cu+Au), gold-gold (Au+Au), and uranium-uranium (U+U) collisions.
The multiplicity densities per participant pair are very similar in all systems studied. Ellipticities vary strongly between collision systems, 
most significantly for central collisions, while fluctuation driven odd moments vary little between systems. 
Event-by-event distributions of eccentricities in mid-central collisions are wider in Cu+Au relative to Au+Au and U+U systems.
An anti-correlation between multiplicity and eccentricity is observed in ultra central U+U collisions which is weaker in the IP-Glasma model than the two-component MC-Glauber model.
In ultra central Au+Au collisions the two models predict opposite signs for the slope of this correlation.
Measurements of elliptic flow as a function of multiplicity in such central events can therefore be used to discriminate between models with qualitatively different particle production mechanisms.

\end{abstract}
  
\maketitle


\section{Introduction}

Understanding how initial state fluctuations influence bulk observables is an important topic in the field of relativistic heavy ion collisions. There are several sources of fluctuations in the initial stages of heavy ion collisions. The dominant ones are the geometric fluctuations of nucleon positions and fluctuations of the impact parameter. 
In collisions of deformed (non-spherical) nuclei the orientations of the nuclei, characterized by four spherical angles, also fluctuate from event to event. For each such configuration of the collision geometry, additional sub-nucleonic fluctuations of color charges lead to fluctuations in the produced gluon fields. 

The combined effect of these fluctuations is reflected in the distributions of global observables such as multiplicity and anisotropic flow. A comparative study of systems with varying initial geometry can provide a better understanding of the relative roles of each source of fluctuations and the details of the  particle generation mechanism. This requires a framework that incorporates all such sources of fluctuations and includes an {\it ab initio} description of multi particle production. 

The IP-Glasma model, based on the Color Glass Condensate (CGC) approach \cite{Gelis:2010nm}, provides such a framework for multi particle production. This model was introduced in \cite{Schenke:2012wb,Schenke:2012hg} and combines the IP-Sat dipole model \cite{Bartels:2002cj,Kowalski:2003hm,Kowalski:2007rw} of lumpy gluon distributions within incoming nuclei, with the classical dynamics of Glasma gluon fields after the nuclear collision \cite{Krasnitz:1999wc,*Krasnitz:2000gz,Lappi:2003bi}. With parameters of the initial lumpy distributions constrained by inclusive and diffractive DIS data from e+p scattering at HERA, the IP-Glasma model can consistently explain the bulk features of global data for various systems like p+p, p+A and A+A over a wide range of energies  \cite{Schenke:2013dpa}. In this paper, we shall employ the model to study a wide range of heavy ion collision systems. In addition to Au+Au collisions at 200 GeV, studied previously in our framework \cite{Schenke:2012hg,Schenke:2013dpa}, we study asymmetric Cu+Au collisions and collisions of highly deformed $^{238}$U nuclei at $200\,{\rm GeV}$ and $193\,{\rm GeV}$, respectively.

The study of collisions of deformed nuclei like $^{238}$U was initially proposed \cite{PBMunzinger:1992, Shuryak:1999by} 
because they promise an additional gain of initial energy density relative to collisions of spherical nuclei. 
A significantly deformed initial geometry at very high energy density for specific orientations of U+U collisions is 
expected to have observable effects on elliptic flow, jet quenching, $J/\psi$ suppression and other observables that characterize 
the properties of the quark gluon plasma (QGP) \cite{Shuryak:1999by, Heinz:2004ir}. 
The interesting configurations with high energy density can be selected in experiments by using a combined cut on the elliptic flow value and the number of spectators for high multiplicity events \cite{Kuhlman:2005ts}. 
Central U+U collisions were further expected to allow for separating signals of the ``chiral magnetic effect (CME)''
from the flow induced background because, for certain configurations, they generate elliptic flow in the absence of a strong magnetic field, as opposed to peripheral collisions of spherical nuclei \cite{Voloshin:2010ut,Bloczynski:2013mca}.

Previous studies \cite{Kuhlman:2006qp, Hirano:2010jg} of U+U collisions within the CGC approach were performed using the Kharzeev-Levin-Nardi (KLN) model~\cite{Kharzeev:2001gp,Kharzeev:2002ei}. 
The single inclusive particle distribution in the KLN model is computed using a $k_T$-factorization approximation. In contrast, the single inclusive particle distribution in the IP-Glasma model is obtained by solving Yang-Mills equations which properly treat momentum modes smaller than the saturation scale. In addition, the IP-Glasma model includes fluctuations of color charges that generate negative binomial $n$-particle distributions. 

The paper is organized as follows. In the next section we briefly discuss the treatment of deformed nuclei in the IP-Glasma model and a two component MC-Glauber model including negative binomial multiplicity fluctuations. In Section \ref{sec_result} we present our results on multiplicity distributions and average eccentricities for different collision systems. We further discuss event-by-event distributions of the ellipticity and its correlation with multiplicity. 
Our conclusions are presented in Section \ref{sec_sum}.

\section{Collisions of deformed nuclei}
\subsection{Collision geometry}
\label{sec_def_ipg}
Collisions of deformed nuclei demand a fully three-dimensional treatment of the system prior to the collision.
In this section, we discuss how the shape and orientation of deformed nuclei are taken into account for the generation of the initial state geometry.
For a review of the theoretical treatment of nuclear shapes, we refer the reader to Ref.\,\cite{Mack1}.

One way to parametrize the densities of deformed nuclei is to modify the Woods-Saxon distribution to read~\cite{Mack2}
 \begin{eqnarray}
 \rho (r, \theta) &=& \frac{\rho_0}{1+ \exp \left( [r-R^\prime(\theta)]/a \right)} \,, \nonumber \\
 {\rm with~}R^\prime(\theta) &=& R\left[ 1+ \beta_2 Y^0_2(\theta) + \beta_4 Y_4^0 (\theta) \right] \,, \nonumber
 \end{eqnarray}
where $\rho_0$ denotes the nucleon density at the center of the nucleus. The spherical harmonic functions $Y^m_l(\theta)$ and the parameters $\beta_2$ and $\beta_4$ account for the deformation from the spherical shape.  This form is by no means unique - such parametrizations in terms of the nuclear density have well known limitations even for spherical nuclei~\cite{Friar}. 
However, as argued in Ref.\,\cite{Mack1}, the data are too few, and the theory uncertainties too large, for a more quantitative and model independent extraction of the shape parameters of deformed nuclei. The discussion in the rest of this paper must therefore be understood to have a systematic uncertainty arising from limitations in our quantitative understanding of the shapes of deformed nuclei. 

\begin{table}
    \begin{tabular}{ | c | c | c | c | c |}
    \hline
    Nucleus& $R [{\rm fm}]$ & $a [{\rm fm}]$ &  $\beta_2$ & $\beta_4$ \\ \hline \hline
    $^{238}$U & 6.81 & 0.55 & 0.28 & 0.093 \\ \hline
    {$^{197}$Au Symmetric }& 6.37 & 0.535 & 0 & 0 \\ \hline
    {$^{197}$Au Deformed }& 6.37 & 0.535 & -0.13 & -0.03 \\ \hline
    $^{63}$Cu & 4.163 & 0.606 & 0 & 0 \\ \hline
    \end{tabular}
    \caption{Parameters for the deformed Woods-Saxon distribution for different nuclei.}
\label{tab:params}
\end{table}

We tabulate in Table \ref{tab:params} parameters of the Woods-Saxon distribution for different nuclei that will be used in this study. 
We use the parameters for $^{238}$U quoted in \cite{Masui:2009qk} because they are currently being investigated in MC-Glauber studies of the STAR collaboration~\cite{paul}.
The \textsc{ampt} calculations of Ref.\,\cite{Haque:2011aa} and the MC-KLN model calculations from Ref.\,\cite{Hirano:2010jg},
 to which we will compare our predictions, also use the same parameters for the deformation for $^{238}$U.
For $^{197}$Au, and $^{63}$Cu, we use the parameters from Ref.\,\cite{DeVries1987495} which include no deformation of these nuclei. Moderate deformation in case of $^{197}$Au is implemented using the parameters from Ref.\,\cite{Moller} as shown in Table \ref{tab:params}.
We emphasize again that since these numbers are extracted from nuclear structure models with a large number of parameters, there is considerable uncertainty in the deformation parameters quoted, and the numbers quoted in the table cannot be considered definitive. For instance, a value of $\beta_2=0.215$ for $^{238}$U is quoted in Ref.\,\cite{Moller} which is 40\% smaller than the value used in \cite{Masui:2009qk} although the same $\beta_4=0.093$ is mentioned in these two references. The value of $\beta_2$ used in \cite{Masui:2009qk} is consistent with results from the experimental measurements of Ref.~\cite{RAMAN20011} which does not quote the value of $\beta_4$. However, we find a variation of $\beta_4$ in the range 0-0.093 to have a negligible effect on the results of this work. 
\begin{figure}[h]
\includegraphics[width=0.4\textwidth]{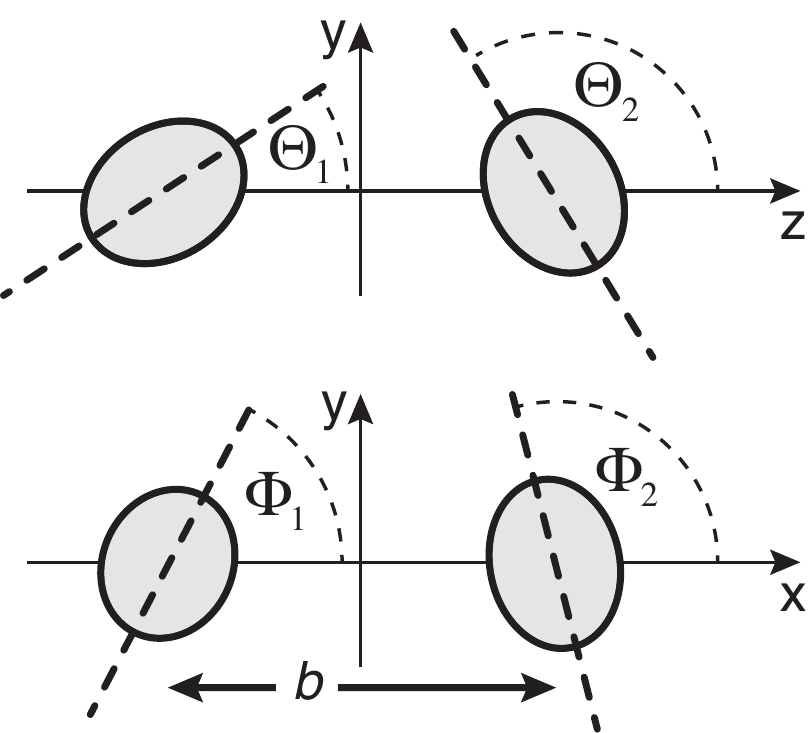}\\
\caption{(Color online) The collision geometry of deformed nuclei. The collision direction is chosen to be along the z-axis and the impact parameter direction is chosen along the x-axis as per convention. The four angles are defined with respect to the major axes of the two colliding nuclei.}
\label{fig_uucoll}
\end{figure}

Nucleons are sampled according to a weight $W(r,\theta)= r^2 \sin(\theta) \rho(r,\theta)$, which includes the Jacobian for the transformation from Cartesian to spherical coordinates. 
As illustrated in Fig.\,\ref{fig_uucoll} for collisions of two deformed nuclei, four angles are used to characterize their orientation.
The two polar angles $\Theta_{1,2}$ denote the orientations of the major axes relative to the collision direction ($z$-axis) and the two azimuthal angles $\Phi_{1,2}$ denote the orientations of the two nuclei in the $x-y$ plane. The impact parameter direction is chosen along the $x$-axis. 

To simulate unpolarized collisions one needs to randomly sample the {polar angles according to the distribution $P(\Theta_{1,2})=\sin(\Theta_{1,2})/2$} between 0 and $\pi$ and the azimuthal angles according to a uniform distribution. The impact parameter $b$ is sampled linearly and the events with no wounded nucleons are rejected. For collisions of uranium nuclei, several special configurations are of particular interest. Configurations with $\Theta_{1,2}=0$ are called tip-tip,  those with $\Theta_{1,2} = \pi/2$ and $\Phi_{1,2}=0$ are called side-side and those with $\Theta_{1,2} = \pi/2$ and $\Phi_{1,2}=\pi/2$  are called body-body\footnote{For body-body collisions the major axes of the two nuclei are aligned with the impact parameter direction.} collisions \cite{Nepali:2007an,Haque:2011aa}.

In this paper, by ``Random U+U'' we refer to unpolarized U+U collisions (averaged over random $\Theta$ and $\Phi$ values) and by ``Au+Au'' we refer to collisions of symmetric Au nuclei. We will present results for both random and tip-tip U+U collisions.
\subsection{MC-Glauber model}
\label{sec_mcg}
A simple model for multi particle production is the two component MC-Glauber model \cite{Kharzeev:2000ph}, wherein the multiplicity is computed from the expression 
\begin{equation}
  \frac{dN}{d\eta} = n_{{\rm pp}} \Big(x N_{\rm coll} + (1-x) \frac{N_{\rm part}}{2}\Big)\,,
\end{equation}
with $n_{\rm pp}$ denoting the average number of charged particles per unit pseudo-rapidity in p+p collisions, $x$  the ``hardness'' scale, 
$N_{\rm coll}$ the number of binary collisions and $N_{\rm part}$ the number of participating nucleons.
When $x=0$, one recovers the expression for the multiplicity in the 
Wounded Nucleon Model~\cite{Bialas:2006kw}. The presence of $N_{\rm coll}$ in this expression introduces a correlation between the multiplicity and the initial shape of the system in the transverse plane at a given $N_{\rm part}$. 

We implement here the MC-Glauber model described in Ref.\,\cite{Masui:2009qk}. For a given impact parameter, a heavy-ion collision is assumed to form $x N_{\rm coll} + (1-x) \frac{N_{\rm part}}{2}$ number of sources, each of which produces particles following a negative-binomial distribution of fixed mean $\bar{n}$ and width $\sim 1/k$:
\begin{equation}\label{eq:nbd}
  P_n^{\rm NB}(\bar{n},k) = \frac{\Gamma(k+n)}{\Gamma(k)\Gamma(n+1)} \frac{\bar{n}^n k^k}{(\bar{n}+k)^{n+k}}\,.
\end{equation}
The mean of the negative binomial distribution $\bar n$ at a given center of mass energy $\sqrt{s_{NN}}$ is obtained from the parametrization of pseudo-rapidity density of charged multiplicity in non-single diffractive $\bar{p}p$ interactions of the form $n_{\bar pp}=2.5-0.25\ln(s_{NN})+0.023\ln^2(s_{NN})$ \cite{Abe:1989td}. Both $k$ and $x$ are free parameters in the MC-Glauber model.

The min-bias multiplicity distribution in this model is obtained by sampling impact parameters according to a linear distribution, sampling nucleon positions and estimating participants and binary collisions using a geometric interpretation of the nucleon-nucleon cross-section to determine the number of sources. One then finally samples $n$ from Eq.\,(\ref{eq:nbd}) for each source and adds up to get the total multiplicity.

\subsection{IP-Glasma model}
\label{sec_ipg}
A detailed description of the IP-Glasma model can be found in \cite{Schenke:2012wb,Schenke:2012hg,Schenke:2013dpa}. Here we use the same model parameters as in \cite{Schenke:2013dpa},
where a detailed discussion of multiplicities and multiplicity distributions from p+p to p/d+A to A+A collisions is given.
Since the IP-Glasma model is based on the CGC framework, $N_{\rm coll}$ and $N_{\rm part}$ do not enter explicitly. For symmetric nuclear collisions, multiplicities are governed by the combination $Q_s^2 S_\perp/\alpha_S$, where $Q_s$ is the saturation scale in one of the two identical nuclei,  $S_\perp$ the transverse overlap area, and $\alpha_S$ the QCD coupling constant. The saturation scale $Q_s^2\propto A^{1/3}$. In general, for deformed nuclei, $A^{1/3}$ needs to be replaced by the number of nucleons along the beam direction--this changes with the orientation of the nuclei. One therefore expects the saturation scale to be larger in tip-tip collisions relative to body-body or side-side collisions. On the other hand, the transverse overlap area is smaller in tip-tip collisions. 
For this reason, it is not transparent how strongly the multiplicity is correlated with the overlap geometry in the IP-Glasma model, and explicit computations are necessary to determine this correlation. The results of these computations will be given in the next section.
\section{Results}

\label{sec_result}

\subsection{Multiplicities}
In the IP-Glasma model, Yang-Mills equations are solved up to time $\tau=0.4\,{\rm fm}/c$, and the transverse Coulomb gauge is fixed, to compute the gluon multiplicity per unit rapidity. A multiplicative factor of 2/3 then converts the gluon multiplicity to the charged particle multiplicity. 

Experimental results for multiplicities are typically presented as a function of $N_{\rm part}$, the number of participant nucleons.
In the IP-Glasma framework, $N_{\rm part}$ does not enter in any of the computations. However, to make comparisons to the experimental data plotted as a function of $N_{\rm part}$, the value of $N_{\rm part}$ is determined geometrically as follows. Two nucleons have an inelastic collision whenever their geometric distance is less than $\sigma_{NN}=42$ mb, the nucleon-nucleon inelastic cross section at the top energy of the Relativistic Heavy Ion Collider (RHIC). We define the total number of nucleons that undergo at least one such inelastic collision to be $N_{\rm part}$.
Note that we neglect the small change of $\sigma_{NN}$ when going from $200\,{\rm GeV}$ to $193\,{\rm GeV}$ in our calculation. 

\begin{figure}[t]
\centering
\includegraphics[angle=-90,width=0.5\textwidth]{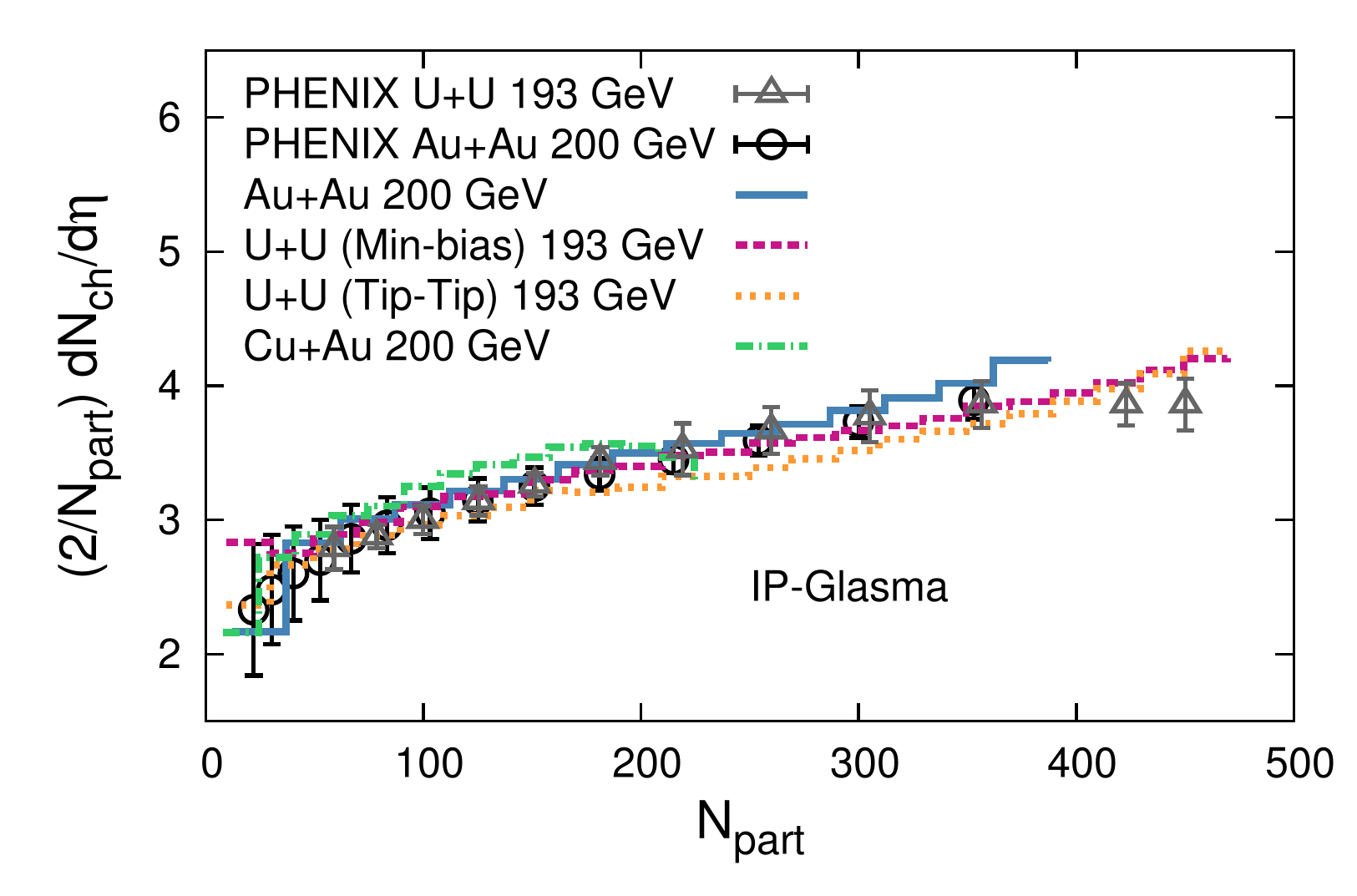}
\caption{(Color online) Centrality dependence of mean multiplicity per participants for different systems. Plotted data points are from Ref.\, \cite{Adler:2004zn,Iordanova:2013jba}.}
\label{fig_multnp}
\end{figure}

The centrality dependence of the mean produced multiplicity density per participant pair $\left(2/N_{\rm part}\right) dN_{\rm ch}/d\eta$ at $\eta=0$ for various systems is shown in Fig.\,\ref{fig_multnp}. Computations of multiplicities in the IP-Glasma model are compared to the available Au+Au data at 200 GeV/nucleon and preliminary results for U+U 193 GeV/nucleon collisions from the PHENIX collaboration~\cite{Iordanova:2013jba}. Fig.~\ref{fig_multnp} also presents the result for Cu+Au collisions, for which no data are as yet available\footnote{For Cu+Au there will be a shift of the rapidity distribution in the laboratory frame by about 0.07 units in the Cu going direction which is negligible and ignored in our calculation.}. The results are very similar for different systems. A weak system size dependence is observed showing that the smaller sized systems Cu+Au and Au+Au produce slightly higher multiplicities per participant compared to U+U collisions. 
Tip-tip U+U collisions produce fewer particles per participant than random collisions for most values of $N_{\rm part}$. Only for the most central events, does the tip-tip configuration produce as many particles per participant as in the random case. 

In the context of the two component model of multiparticle production discussed above, tip-tip configurations should produce higher multiplicities per participant pair since more nucleons are aligned along the beam direction, corresponding to a larger number of binary collisions for a given $N_{\rm part}$. This was indeed found in calculations shown in \cite{Kuhlman:2005ts}, as well as in a computation in the \textsc{ampt} model \cite{Haque:2011aa}.
The IP-Glasma model does not show this behavior. Likewise, the KLN model~\cite{Kharzeev:2001gp,Kharzeev:2002ei} yields smaller differences in multiplicity between different orientations of the uranium nuclei relative to computations in the two-component MC-Glauber model \cite{Kuhlman:2006qp}.

For Cu+Au collisions we see an interesting centrality dependence of multiplicity which is not seen in case of other systems. As shown in Fig.~\ref{fig_multnp}, the centrality dependence of $\left(2/N_{\rm part}\right) dN_{\rm ch}/d\eta$ for Cu+Au flattens out above $N_{\rm part}>180$. One possible interpretation could be that the Cu nucleus is completely surrounded by the Au nucleus for most central Cu+Au collisions. In this case, the minimum saturation scale among the two nuclei that controls the multiplicity does not grow fast enough with further increase of $N_{\rm part}$.

\begin{figure}[h]
\centering
\includegraphics[width=0.5\textwidth]{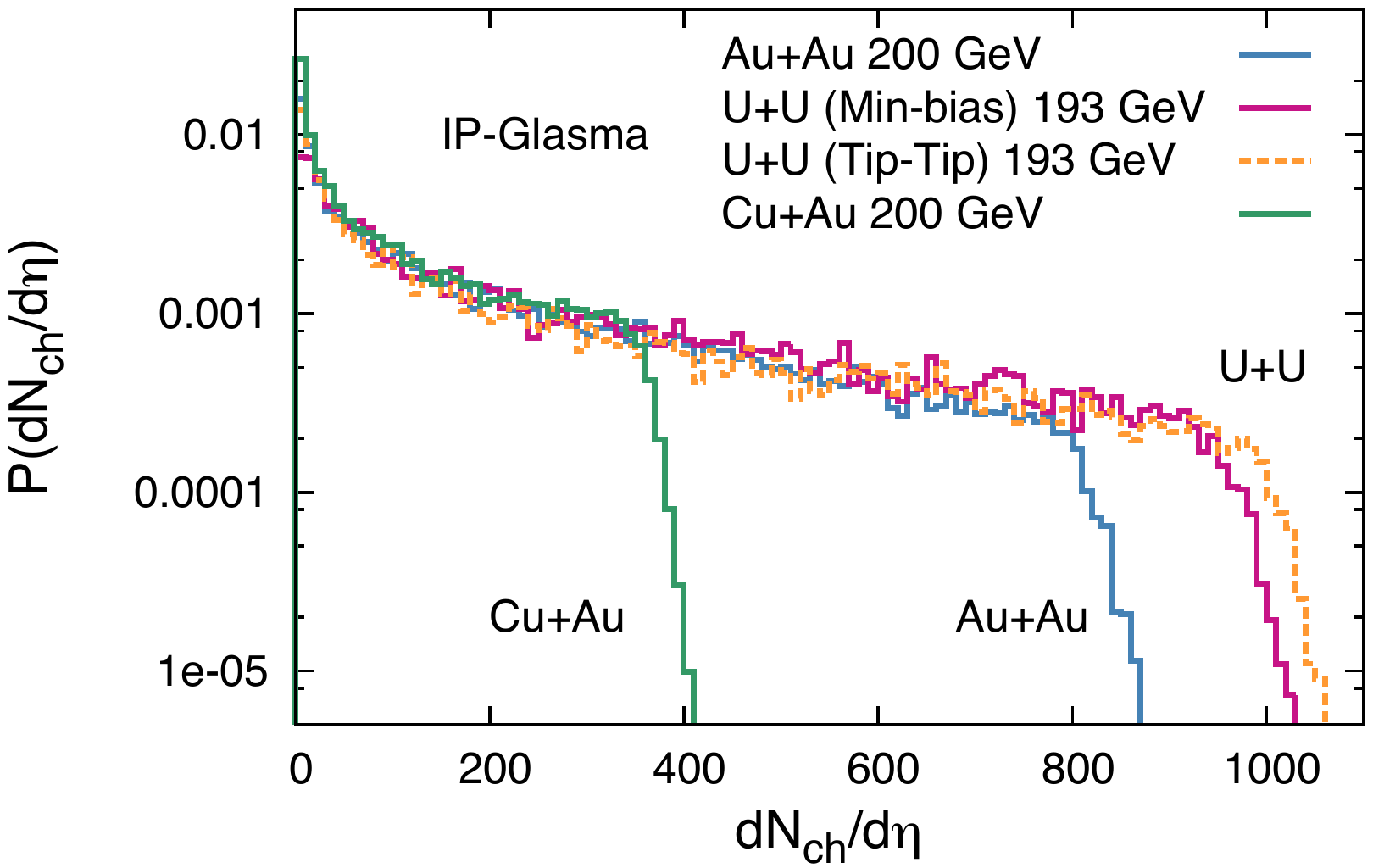}
\caption{(Color online) Probability distributions of charge particle multiplicity for different systems.}
\label{fig_multdist}
\end{figure}

As noted, the IP-Glasma model naturally produces negative binomial multiplicity distributions for a fixed collision geometry \cite{Schenke:2012hg}. The multiplicity distribution in A+A collisions is a convolution of such NBDs for each initial configuration. To compute the probability distribution of the multiplicity, we sample collisions over a wide range of impact parameters (that follow a linear distribution) and reject events that do not produce any wounded nucleons. This approach leads to results equivalent to weighting events with an eikonal weight function \cite{Schenke:2012hg}.
The multiplicity distributions for different systems are shown in Fig.\,\ref{fig_multdist}. 
We see an approximately $20\%$ larger maximal multiplicity for U+U collisions compared to Au+Au collisions.
The difference in the maximal multiplicity between tip-tip and random U+U collisions is small ($\le 5\%$). 

The shape of the multiplicity distribution for random U+U collisions looks similar to the Glauber model prediction of Ref.\,\,\cite{Kuhlman:2005ts} and is slightly wider than the KLN model prediction of \cite{Kuhlman:2006qp}. It should be noted however that for these MC-Glauber and KLN computations Gaussian event-by-event fluctuations (as opposed to negative binomial distributions) of multiplicity were introduced by hand. The free parameters controlling the widths of the Gaussian distributions were fixed using a fit to Au+Au data. 

In our implementation of the MC-Glauber model introduced in section \ref{sec_mcg}, the parameter $k$ that controls the width of the negative binomial distribution and the parameter $x$ are extracted by fitting the min-bias multiplicity distribution obtained from the IP-Glasma model. We find that $k=8$ and $x=0.135$ provide approximate agreement with the IP-Glasma min-bias multiplicity distribution for both Au+Au collisions at 200 GeV and U+U collisions at 193 GeV. In the following section, we use this tuned version of the MC-Glauber model for further ``apples-to-apples" comparisons to IP-Glasma model predictions.

Studies of global observables in Cu+Au collisions were previously performed using \textsc{ampt} simulations \cite{Chen:2005zy, Haque:2011ti} and MC-Glauber+hydrodynamic simulations \cite{Bozek:2012hy}. These references do not discuss multiplicity fluctuations and only quote the average multiplicity for selected centralities. To compare our results with such predictions, we compute multiplicity in different centrality classes using the multiplicity distributions shown in Fig.\,\ref{fig_multdist}. However it must be noted that the centrality selections in the above mentioned models are done using participant or impact parameter distributions. For min-bias Cu+Au collisions, we find a multiplicity $dN_{\rm ch}/d\eta$ of $\sim 90$ at $\eta=0$, slightly lower than the \textsc{ampt} prediction of \cite{Chen:2005zy}. For $0-5\%$ and $20-30\%$ centralities, we find  $dN_{\rm ch}/d\eta =340 \pm 18$ and $150 \pm 17$ respectively which are very close to the hydrodynamic model prediction of  \cite{Bozek:2012hy}. 
\subsection{Initial energy distributions and collision geometry}
\begin{figure}[b]
\includegraphics[angle=-0,width=0.235\textwidth]{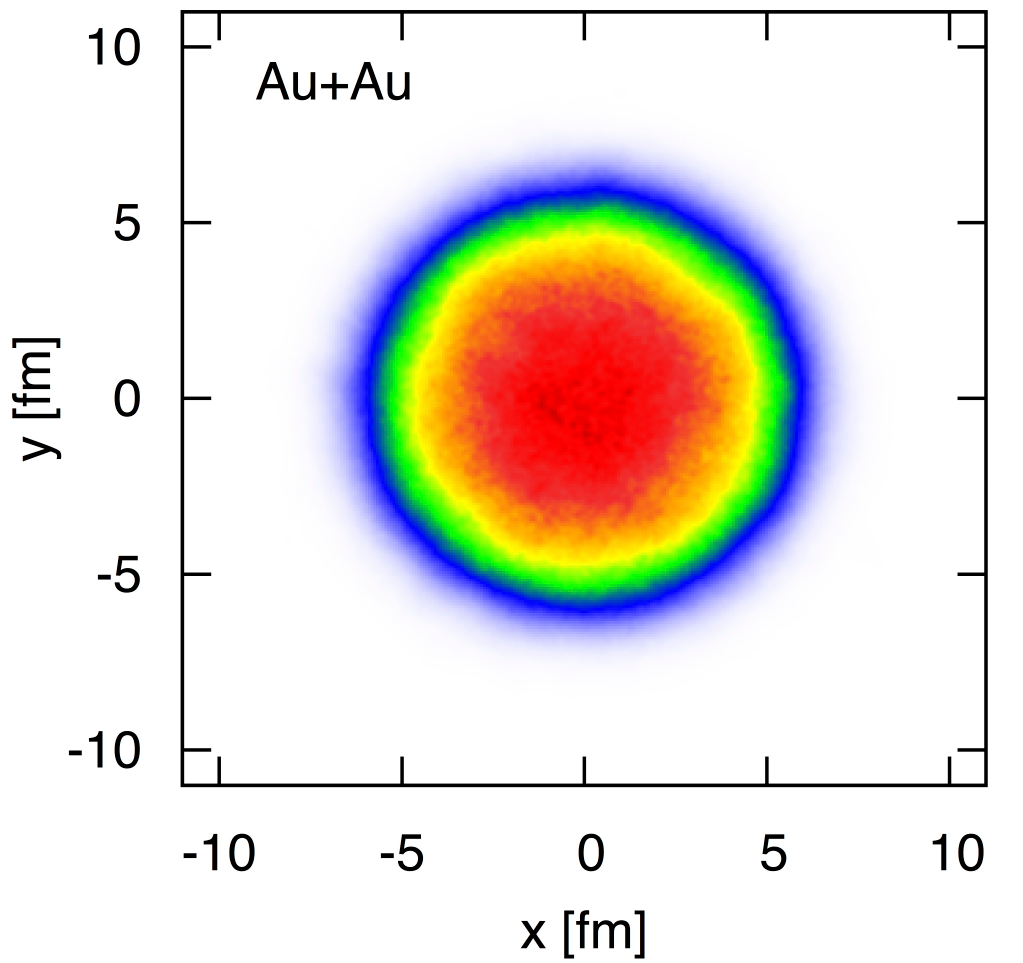}
\includegraphics[angle=-0,width=0.235\textwidth]{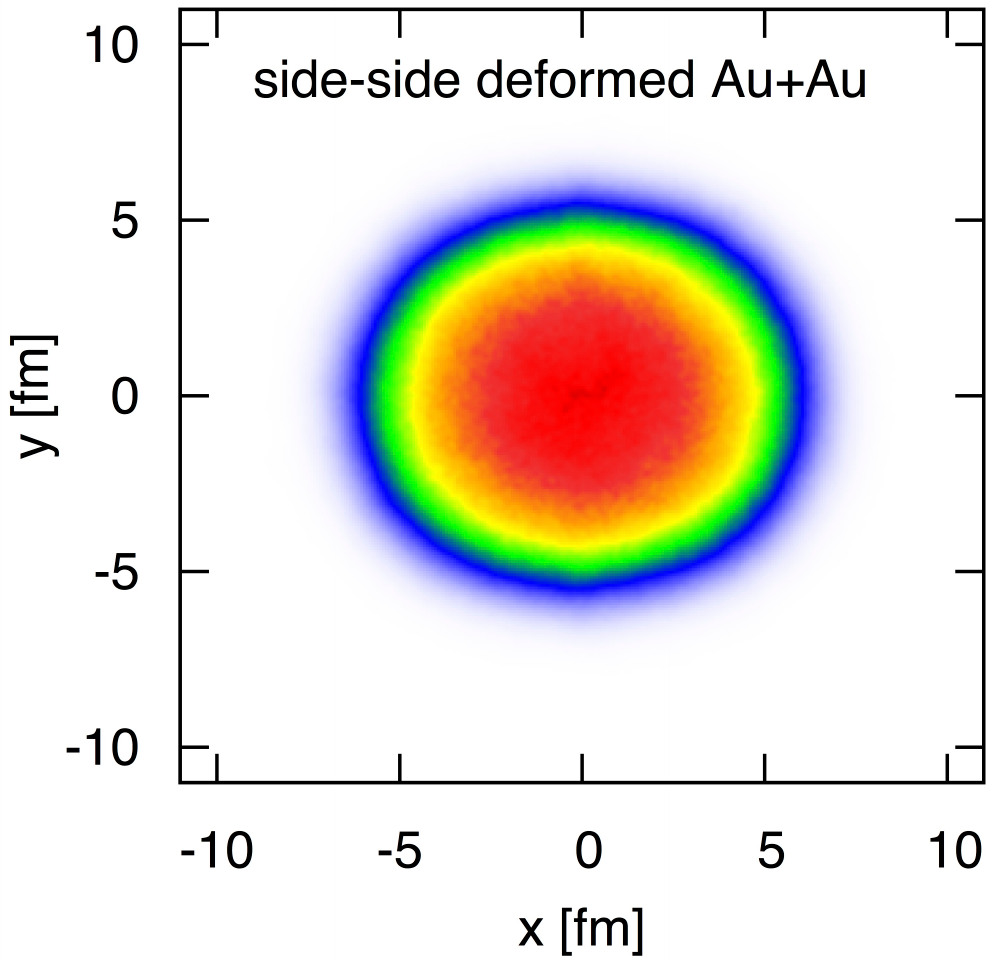}
\includegraphics[angle=-0,width=0.23\textwidth]{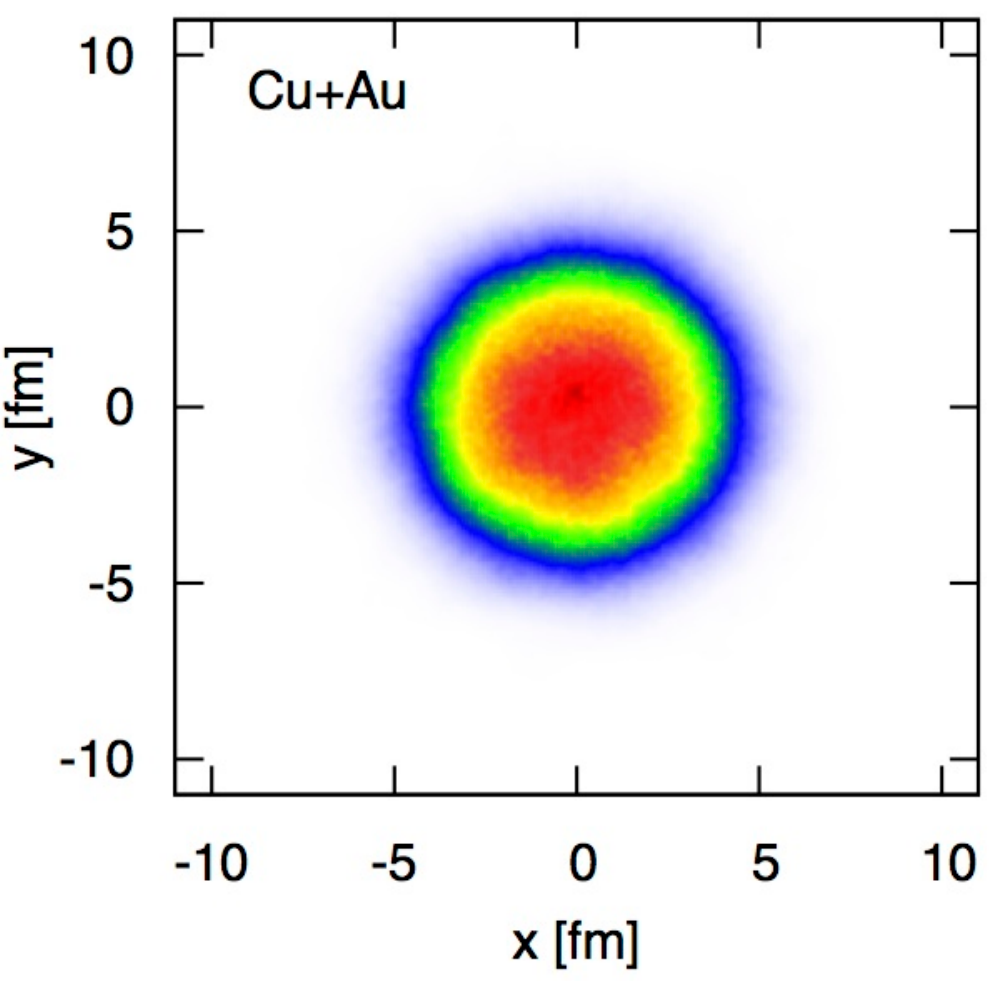}
\caption{(Color online) Energy density (in arbitrary units, increasing density from blue to red) distribution averaged over 1000 IP-Glasma events in the transverse plane at the initial time for Au+Au (upper panel) and Cu+Au (lower panel) collisions at zero impact parameter.} 
\label{fig_eden1}
\end{figure}
\begin{figure*}[t]
\centering
\scalebox{0.5}
{
\includegraphics[angle=-0,width=0.5\textwidth]{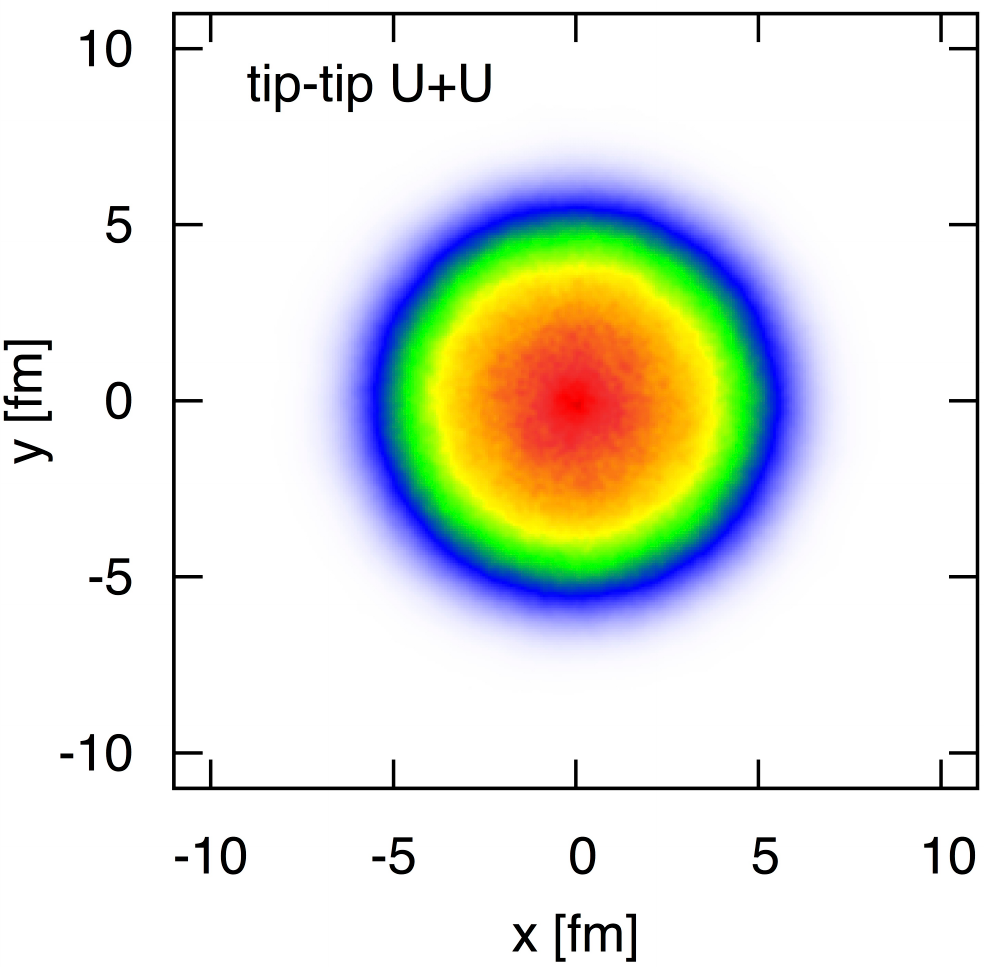}
\includegraphics[angle=-0,width=0.5\textwidth]{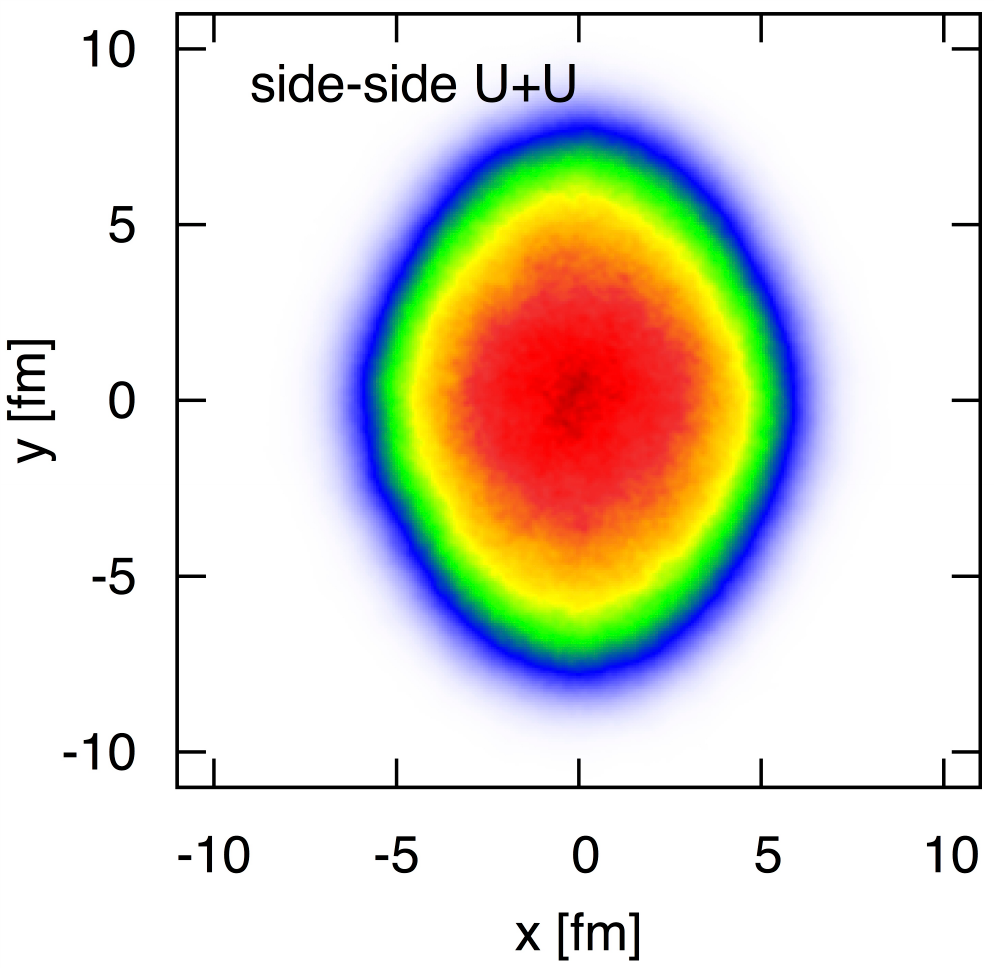}
\includegraphics[angle=-0,width=0.5\textwidth]{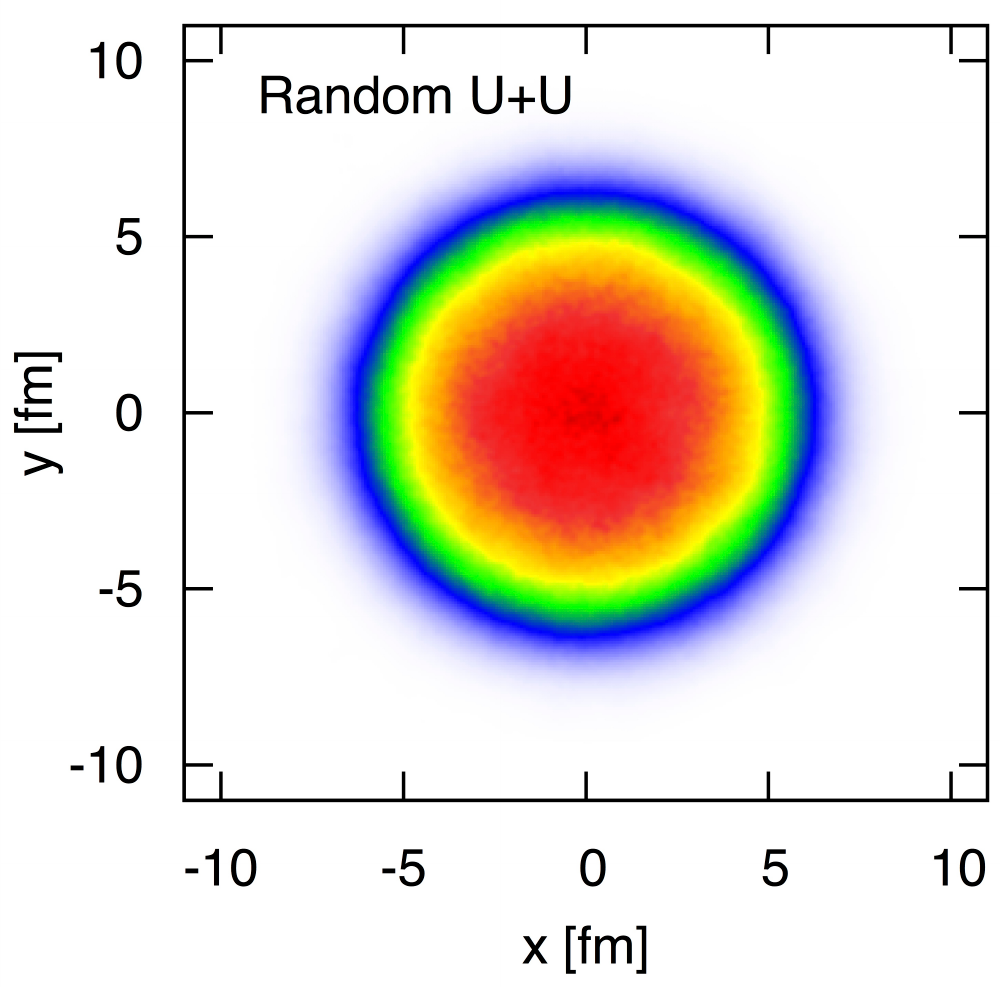}
\includegraphics[angle=-0,width=0.5\textwidth]{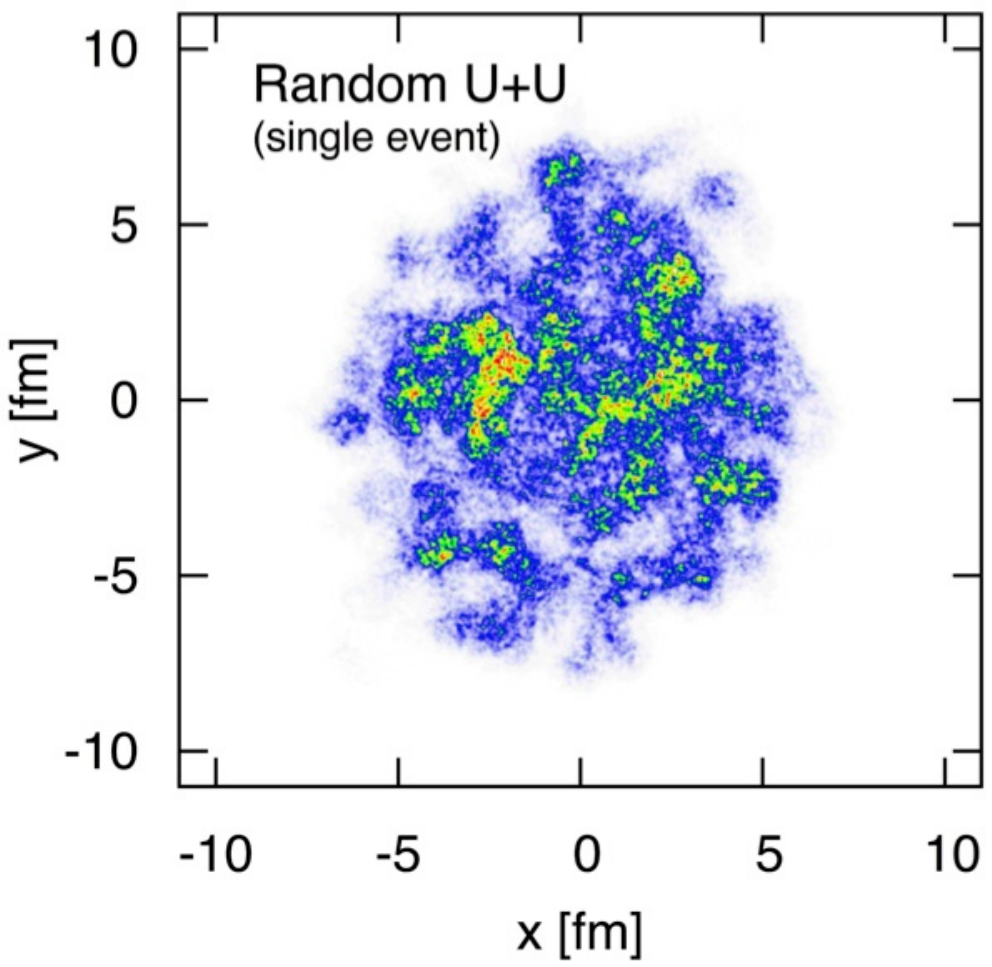}
}
\caption{(Color online) Distribution of energy density (in arbitrary units, increasing density from blue to red) in the transverse plane at the initial time averaged over 1000 IP-Glasma events for different configurations of $U+U$ collisions at $b=0$. From left to right distributions are shown for tip-tip, side-side, and random configurations of U+U collisions. The right most panel shows a random single event distribution from the IP-Glasma model.}
\label{fig_eden2}

\end{figure*}

Average initial energy density distributions from IP-Glasma model in the transverse plane for zero impact parameter ($b=0$) in different systems are shown in Figs.\,\ref{fig_eden1} and \ref{fig_eden2}. Averages are taken over 1000 events. The upper panel of Fig.\,\ref{fig_eden1} shows the symmetric Au+Au collisions and the side-side configuration of the deformed Au+Au collisions.
Comparing the lower and the upper panels of Fig.\,\ref{fig_eden1} one can see that for Cu+Au collisions the system size is dominated by the smaller size of the Cu nucleus.

 Event averaged energy density distributions for different configurations of U+U collisions are shown in Fig.\,\ref{fig_eden2}. The tip-tip configuration produces the smallest system, comparable in size to central Au+Au collisions. The significant prolate deformation of uranium nuclei is visible in case of side-side collisions. The event averaged size of unpolarized U+U collisions is shown for comparison, as is the spatial distribution of energy density in a single event. 
\subsection{Eccentricities}
\begin{figure}[thb]
\centering
\includegraphics[angle=-90,width=0.5\textwidth]{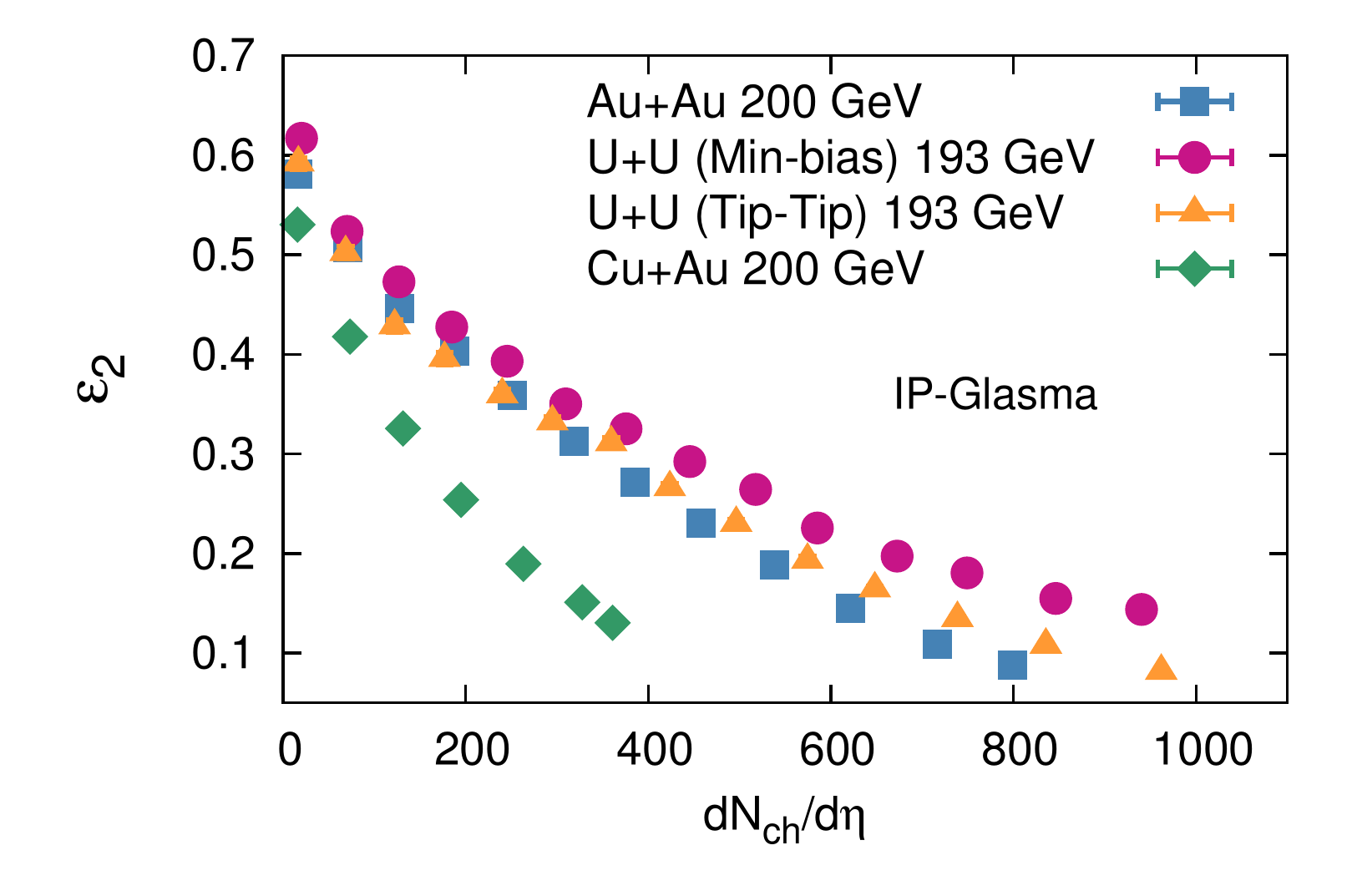}
\caption{(Color online) Multiplicity dependence of initial ellipticity for different systems.}
\label{fig_ecc}
\end{figure}
\begin{figure}[htb]
\centering
\includegraphics[angle=-90,width=0.5\textwidth]{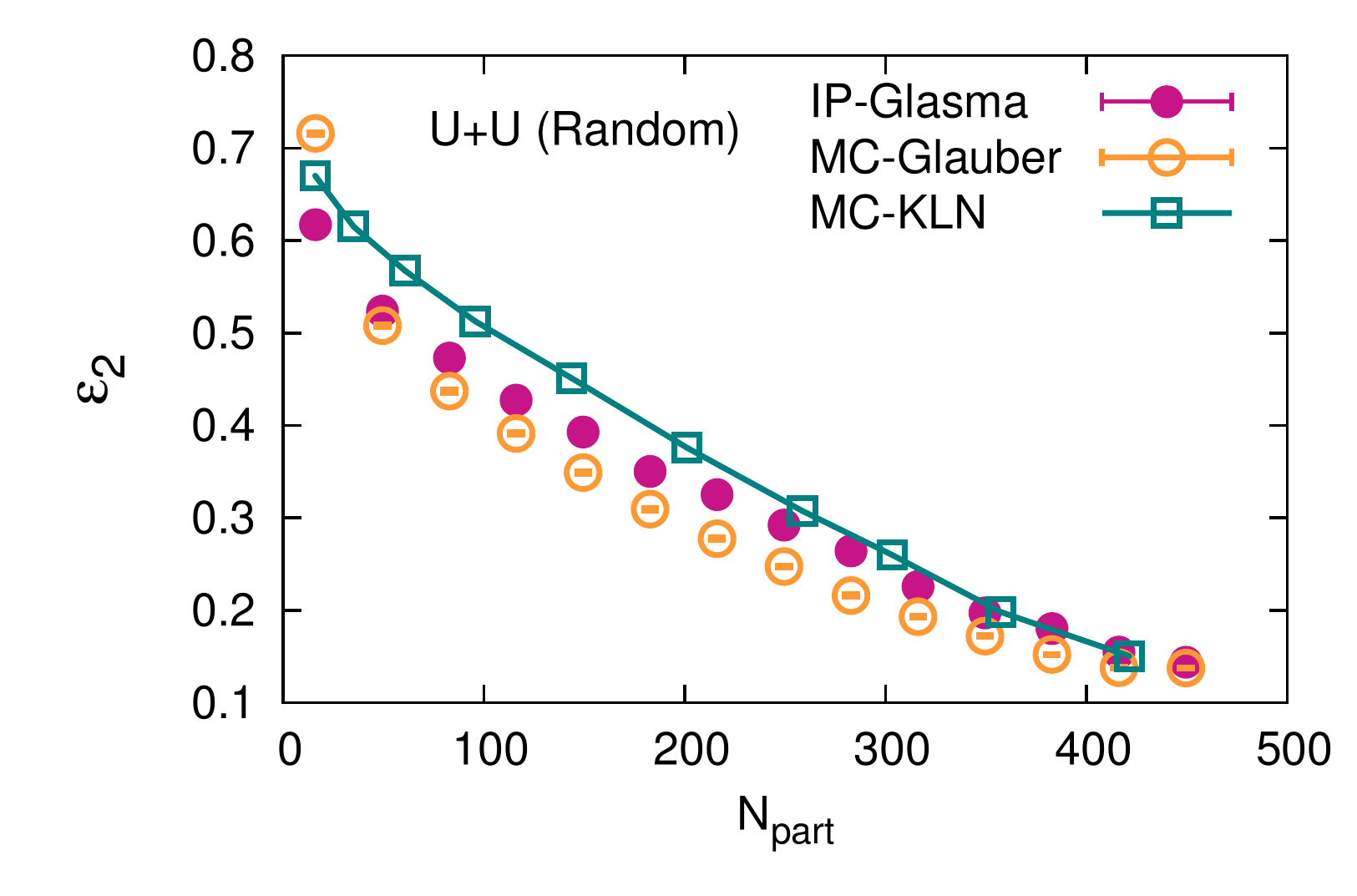}
\caption{(Color online) Comparison of $\varepsilon_2$ for random U+U collisions from different models. The MC-KLN result is from Ref.\,\cite{Hirano:2010jg}. }
\label{fig_modelUU}
\end{figure}

The $n$-th order spatial eccentricity that characterizes the initial state geometry is defined as
\begin{equation}
  \varepsilon_n = \frac{\sqrt{\langle r^n \cos(n\phi)\rangle^2+\langle r^n \sin(n\phi)\rangle^2}}{\langle r^n \rangle}\,
\end{equation}
Here $\left< \cdot \right>$ is the energy density $\epsilon(r,\phi,\tau)$ weighted average. 
To eliminate noise in the computation of eccentricities, we only include cells in which the energy density is greater than $\epsilon_{\rm min}=\Lambda_{\rm QCD}^4$, where $\Lambda_{\rm QCD}$ is chosen to be 200 MeV. The effect due to variation of $\epsilon_{\rm min}$ was previously studied in Ref. \cite{Bzdak:2013zma}.
We show results for eccentricities evaluated at the initial time after the collision.
\begin{figure}
\centering
\includegraphics[angle=-90,width=0.5\textwidth]{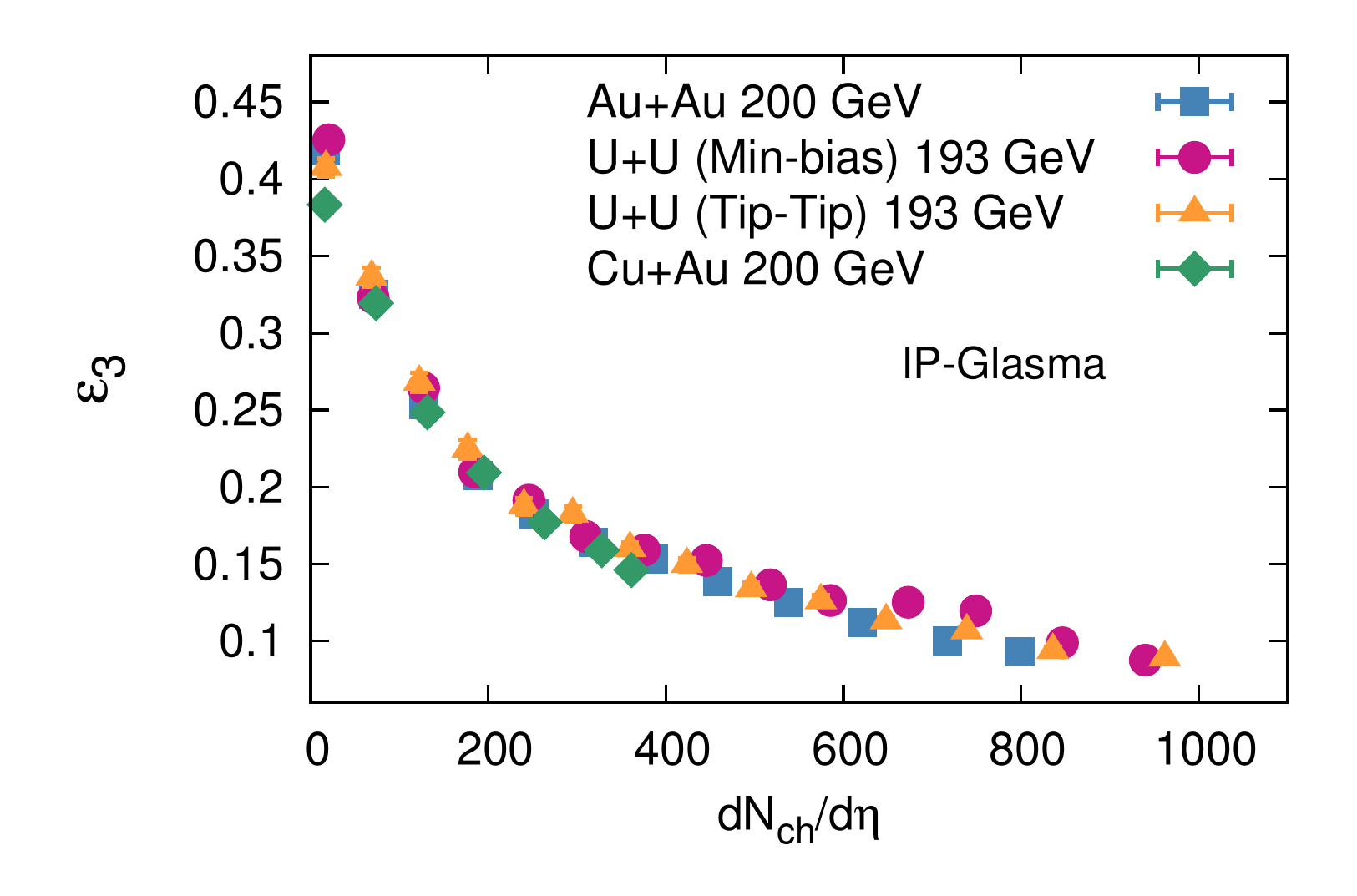}
\caption{(Color online) Variation of the initial triangularity with produced charged particle multiplicity for different collision systems. }
\label{fig_trgnp}
\end{figure}

Fig.\,\ref{fig_ecc} shows the initial ellipticity $\varepsilon_2$ as a function\footnote{Results for $\varepsilon_2$ as a function of $N_{\rm part}$ show a very similar behavior to those presented as a function of $dN_{\rm ch}/d\eta$ for all systems.} of $dN_{\rm ch}/d\eta$.
$\varepsilon_2$ is larger for random U+U collisions than tip-tip configurations because the former include side-side events that give rise to large values of $\varepsilon_2$ 
(see Fig.\,\ref{fig_eden2}).  For the highest multiplicity, this difference amounts to a factor of approximately 1.5. %

The multiplicity dependence of $\varepsilon_2$ is flatter for random U+U collisions than tip-tip U+U collisions above $dN_{\rm ch}/d\eta \gtrsim 700$. The shape of the tip-tip U+U curve is very similar to that of the Au+Au curve except for an overall shift due to the larger number of nucleons in the uranium nucleus. 
The values of $\varepsilon_2$ merge for Au+Au and U+U collisions towards peripheral bins.  

$\varepsilon_2$ shows a significant difference in terms of both magnitude and the trend with $dN_{\rm ch}/d\eta$ in case of Cu+Au collisions. At low $dN_{\rm ch}/d\eta$ the value of $\varepsilon_2$ in Cu+Au is comparable to other systems. However it falls off much faster with $dN_{\rm ch}/d\eta$.

In Fig.\,\ref{fig_modelUU} we compare the ellipticity in random U+U collisions from different models. 
The MC-Glauber result is obtained using the model described in Section \ref{sec_mcg}. Here, the ellipticity is computed by averaging over all participant nucleon positions defined by the nucleon centers. The MC-KLN model calculation is taken from Ref.\,\cite{Hirano:2010jg}. The MC-KLN model produces the largest $\varepsilon_2$ over a wide range of $N_{\rm part}$. The MC-Glauber $\varepsilon_2$ increases rapidly at low $N_{\rm part}$ to reach the limiting value of $\varepsilon_2=1$ at $N_{\rm part}=2$.

The triangularity $\varepsilon_3$ for different collision systems is shown as a function of $dN_{\rm ch}/d\eta$ in Fig.\,\ref{fig_trgnp}.  $\varepsilon_3$ values for all systems coincide over the entire range of $dN_{\rm ch}/d\eta$ which is a striking reflection of the fact that $\varepsilon_3$ is sensitive only to fluctuations, not to the details of the average geometries. 
A similar behavior of $\varepsilon_3$ was also seen in the \textsc{ampt} model calculations of\,\cite{Haque:2011aa,Haque:2011ti}.

\begin{figure}[h]
\centering
\includegraphics[angle=-90,width=0.5\textwidth]{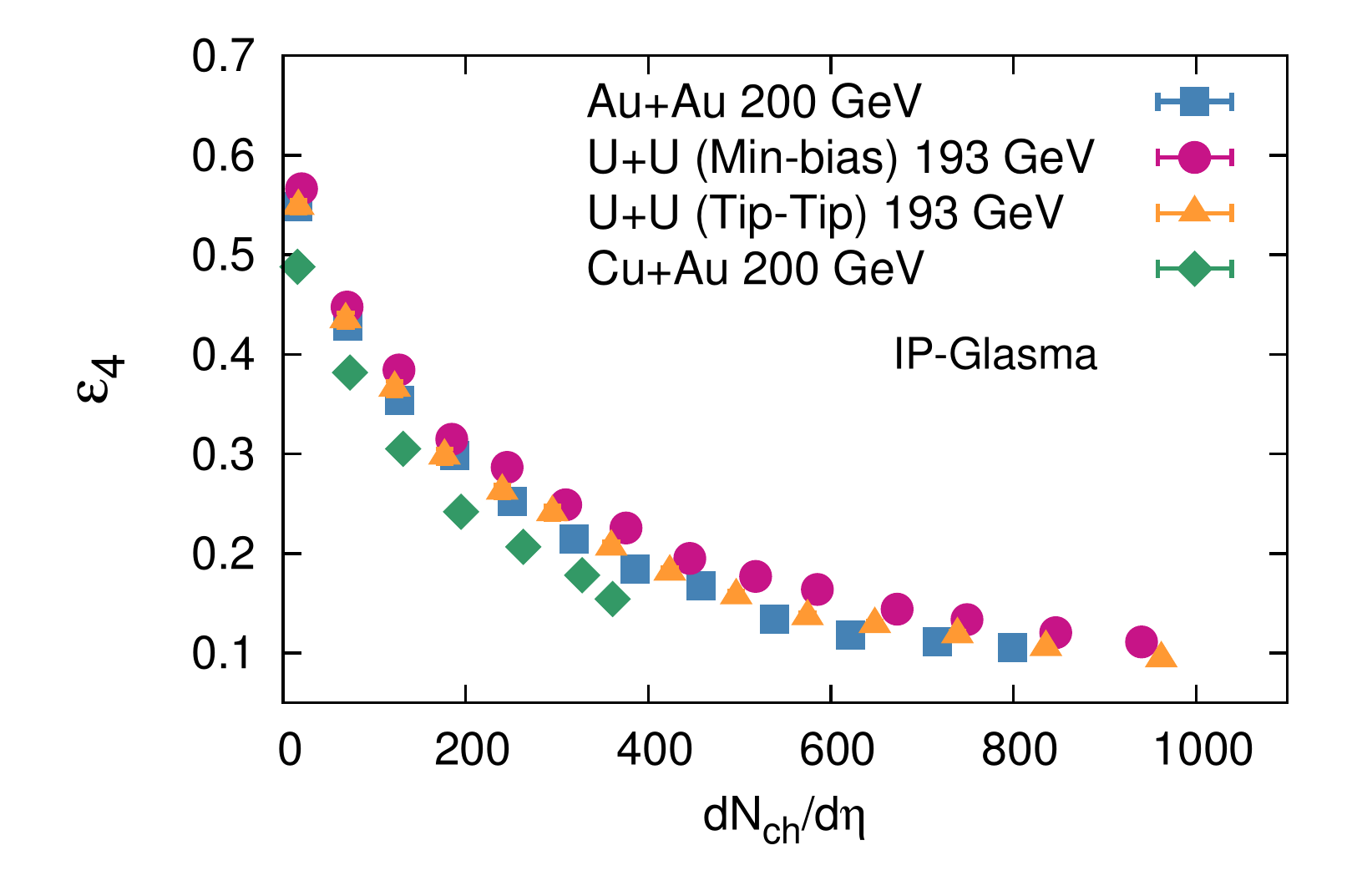}
\includegraphics[angle=-90,width=0.5\textwidth]{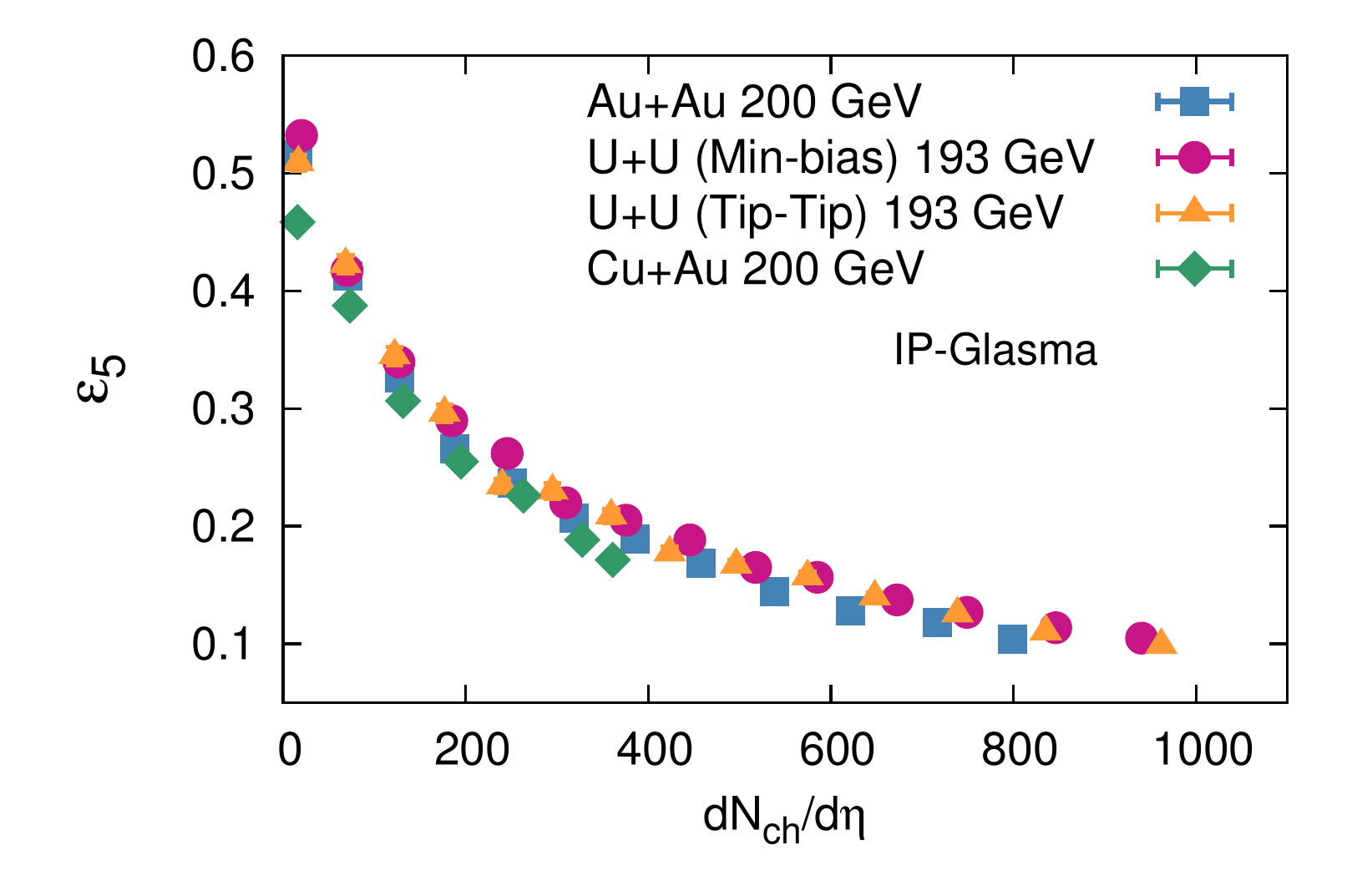}
\caption{(Color online) Higher order of eccentricities $\varepsilon_4$ (upper panel) and $\varepsilon_5$ (lower panel) for different systems plotted as a function of produced charged particle multiplicity.}
\label{fig_ecc4np}
\end{figure}
Higher order moments of eccentricities as functions of  $dN_{\rm ch}/d\eta$ are shown in Fig.\,\ref{fig_ecc4np}. $\varepsilon_4$ for U+U and Au+Au coincide over the entire range of $dN_{\rm ch}/d\eta$. For Cu+Au $\varepsilon_4$ is slightly lower. As for $\varepsilon_3$, the $dN_{\rm ch}/d\eta$  dependence of $\varepsilon_5$ is very similar in all systems.
\subsection{Event-by-event fluctuations of ellipticities}
\label{sec_ebye}
\begin{figure}[h]
\centering
\includegraphics[trim=0cm 0cm 0cm 1cm,width=0.48\textwidth]{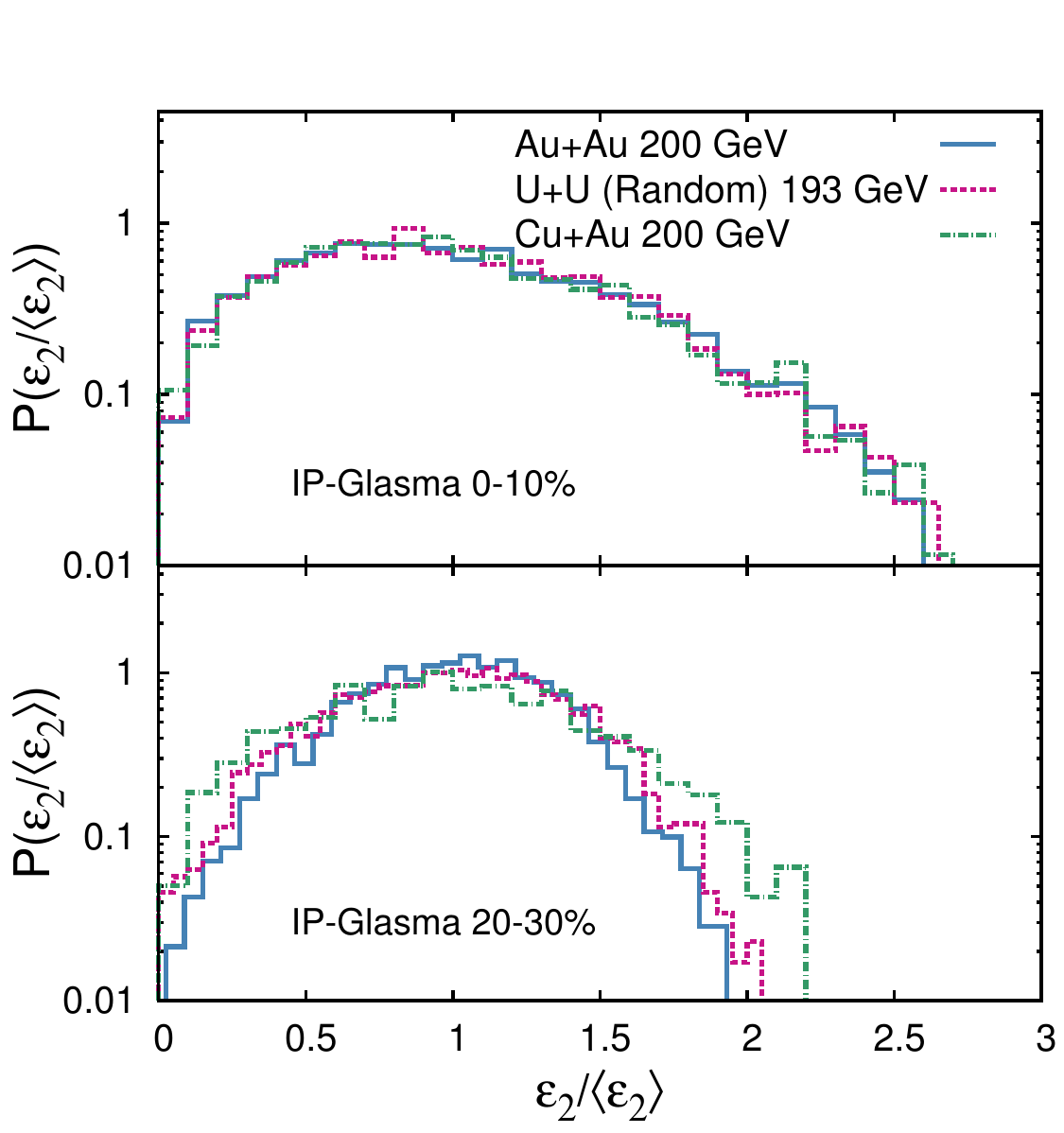}
\caption{(Color online) Probability distributions of scaled initial ellipticity for different systems. Distributions are shown for two centrality classes $0-10\%$ and $20-30\%$.}
\label{fig_eccdist}
\end{figure}
Event-by-event fluctuations of ellipticities are sensitive to the details of initial state fluctuations and provide a good estimate of $v_2$ fluctuations~\cite{Schenke:2013aza, Renk:2014jja}. 
Our computations in Refs.\,\cite{Gale:2012in, Gale:2012rq} demonstrated that the distributions of scaled eccentricities $\varepsilon_n/\left< \varepsilon_n \right>$ using the IP-Glasma model provide a very good description of the experimental $v_n/\langle v_n\rangle$ distributions measured by the ATLAS collaboration \cite{Aad:2013xma}. In \cite{Schenke:2013aza} we extended the calculations to 10 centrality bins in the range of 0-50\% and obtained good agreement with ATLAS data for all centralities. Deviations from the experimental data are only found in the large $\varepsilon_n$ ($v_n$) tails of the distributions where non-linear effects of the hydrodynamic evolution become important \cite{Gale:2012rq}.

The ATLAS collaboration has demonstrated that both MC-Glauber and MC-KLN models are unable to explain the data across the full range of centrality \cite{Jia:2012ve}. These results indicate that event-by-event distributions of $v_n$ can be very powerful observables to discriminate between different models of initial conditions.

Thus far no attempt has been made to measure the $v_n$ distributions at RHIC. The first cumulant of the $v_2$ distribution has been measured by the STAR collaboration~\cite{Agakishiev:2011eq}. However, this does not entirely constrain the complete distribution. Measurements of $v_n$ distributions at RHIC are important to verify the applicability of different initial state models at lower energies. The lower multiplicities at RHIC compared to the LHC make the experimental extraction of $v_n$ distributions more challenging (see \cite{Aad:2013xma}). However, the analysis should be feasible at RHIC for centralities (central or semi-central U+U or Au+Au collisions) in which the multiplicities are comparable to mid-central or peripheral Pb+Pb collisions at LHC.

As done in Ref.\,\cite{Schenke:2013aza} for Pb+Pb collisions at LHC energies, we compute the event-by-event distributions of the scaled ellipticity for different systems at the highest RHIC energy. 
These predictions can be compared with the experimental distributions of scaled $v_2/\left<v_2\right>$. 
Results for two centrality classes $0\!-\!10\%$ and $20\!-\!30\%$ are shown in Fig.\,\ref{fig_eccdist}. For more peripheral bins, the $\varepsilon_2/\left< \varepsilon_2 \right>$ distribution may not provide a good prediction of the corresponding $v_2/\left<v_2\right>$ distribution as discussed above. The computation of $v_2/\left<v_2\right>$ distributions at RHIC in our framework  combined with viscous hydrodynamic simulations is left for future work.

As shown in Fig.\,\ref{fig_eccdist} the scaled $\varepsilon_2$ distributions for $0\!-\!10\%$ events are very similar for different systems.
This indicates that the widths of the ellipticity distributions are proportional to the corresponding mean values of ellipticity for central events. 
In the $20\!-\!30\%$ centrality bin differences between different collision systems are more prominent. In particular Cu+Au collisions have a noticably wider distribution. 
\begin{figure}[h]
\centering
\includegraphics[trim=0cm 0cm 0cm 1cm,width=0.48\textwidth]{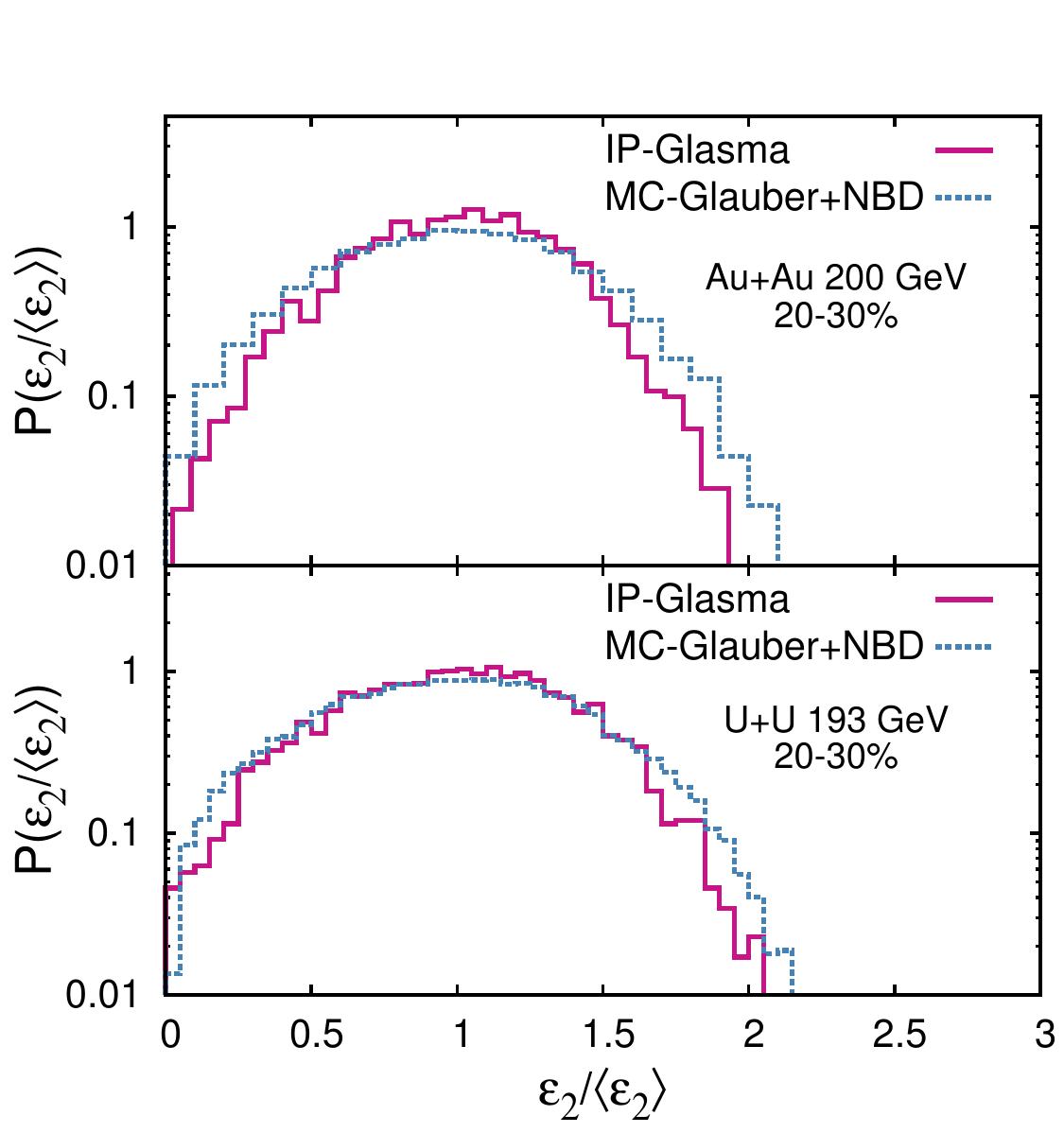}
\caption{(Color online) Probability distributions of the scaled initial ellipticity for symmetric Au+Au collisions and random U+U collisions for $20-30\%$. Distributions are compared to MC-Glauber results.}
\label{fig_eccdist_mc}
\end{figure}

In Fig.\,\ref{fig_eccdist_mc} we compare the IP-Glasma model results with the $\varepsilon_n$ distributions from the two-component MC-Glauber model for Au+Au and U+U collisions. The centrality selections in the MC-Glauber model is done using the corresponding min-bias multiplicity distribution. As shown, the widths of the $\varepsilon_n$ distributions are larger in case of the MC-Glauber model for $20-30\%$ central collisions. The difference is more prominent in case of Au+Au collisions. For $0-10\%$ central collisions we find this difference to be negligible. 
\subsection{Correlation of ellipticity and multiplicity in ultra-central events}\label{sec_corr}
\begin{figure}[htb]
\centering
\includegraphics[angle=-90,width=0.48\textwidth]{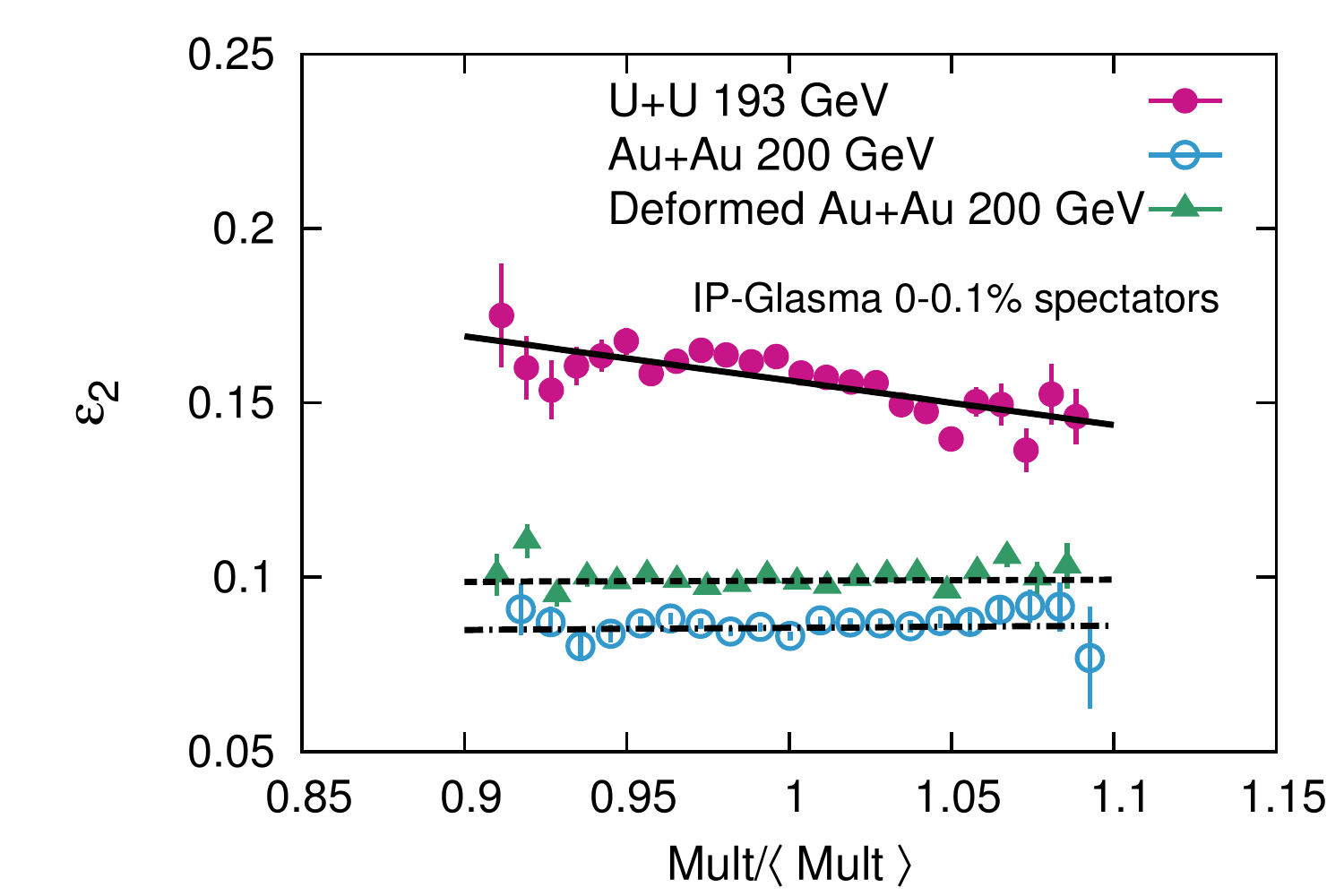}
\caption{(Color online) Variation of initial ellipticity with scaled multiplicity for different systems. Results are shown for events with $0.1\%$ fraction spectator nucleons. The black lines are linear fits to IP-Glasma points.}
\label{fig_eccmult1}
\end{figure}
\begin{figure}[htb]
\centering
\includegraphics[width=0.48\textwidth]{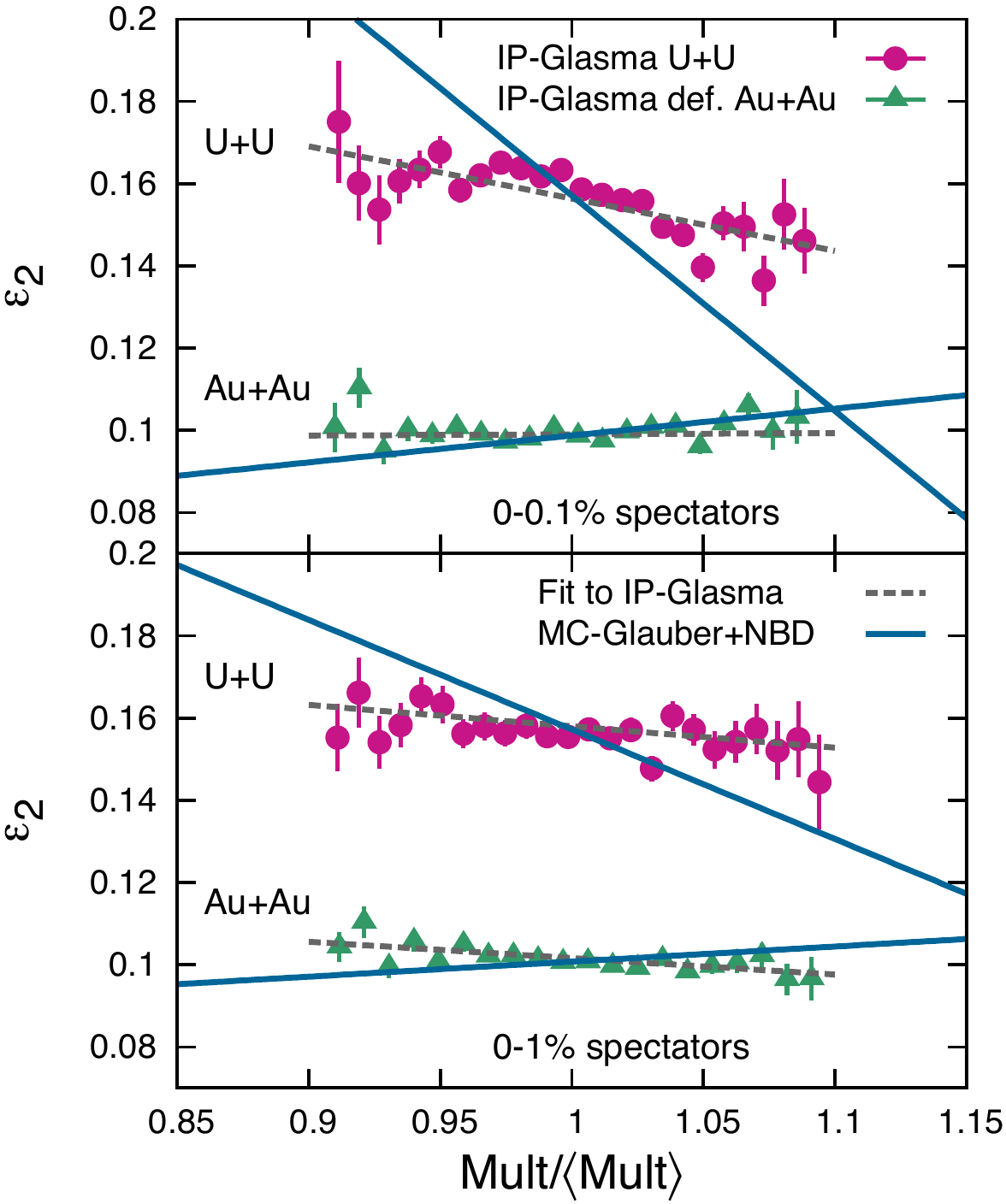}
\caption{(Color online) Variation of initial ellipticity with the scaled multiplicity in case of deformed U+U and deformed Au+Au collisions. Results are shown for events with $0\!-\!0.1\%$ and $0\!-\!1\%$ fraction spectator nucleons. We further show linear fits to the IP-Glasma results.}
\label{fig_eccmult2}
\end{figure}

To analyze the correlation between the multiplicity and the overlap geometry, we determine $\varepsilon_2$ as a function of the scaled multiplicity in ultra-central events.
These events are determined by using strong cuts on the distribution of the spectator nucleons, as was suggested in \cite{Kuhlman:2005ts}. We make predictions for the $0\!-\!0.1\%$ and $0\!-\!1\%$ most central events.\footnote{For Au+Au collisions $0\!-\!0.1\%$ and $0\!-\!1\%$ events correspond to a total of maximally 8 and 16 spectator neutrons, respectively. For U+U the same centrality bins correspond to 12 and 24 spectator neutrons, respectively. These cuts are obtained using MC-Glauber simulations by the STAR collaboration~\cite{hwang}. The number of neutrons among all spectator nucleons are sampled from a binomial distribution with probability (1-Z/A), where Z and A are atomic and mass numbers of the nucleus.}
For U+U collisions specifically, these ultra-central events are either (almost) full overlap tip-tip, side-side, or body-body collisions. 

In Figs.\,\ref{fig_eccmult1} and \ref{fig_eccmult2} we show IP-Glasma model results for the eccentricity $\epsilon_2$ as a function of the multiplicity (scaled by the average multiplicity in the $0\!-\!0.1\%$ and $0\!-\!1\%$ bin, respectively).
In Fig.\,\ref{fig_eccmult1}, we compare results for U+U collisions, deformed (oblate) Au+Au collisions, and Au+Au collisions with the deformation parameters set to zero in the $0\!-\!0.1\%$ centrality bin. The results shown are extracted from simulations of 20,000 events. For U+U collisions, we notice a distinct anti-correlation between eccentricity and multiplicity. The slopes of both deformed and spherical Au+Au collisions are consistent with zero within the statistical uncertainties.

In Fig.\,\ref{fig_eccmult2}, we compare the IP-Glasma model results with those from the two-component MC-Glauber model. The top and bottom panel of  Fig.\,\ref{fig_eccmult2} show the $0\!-\!0.1\%$ and $0\!-\!1\%$ centrality bins, respectively. As mentioned in Section \ref{sec_mcg}, to allow for an ``apples-to-apples'' comparison, the parameters of the Glauber model are adjusted to ensure that the corresponding min-bias multiplicity distribution is in approximate agreement with the IP-Glasma multiplicity distribution shown in Fig.~\ref{fig_multdist}. The comparison shows that there is a much stronger anti-correlation between eccentricity and multiplicity for U+U collisions in the MC-Glauber model relative to the IP-Glasma model. For (oblate) deformed Au+Au collisions, the two models give qualitatively different results. 

In the $0\!-\!1\%$ centrality bin, the IP-Glasma model yields a slope of $-0.03\pm0.01$, while the MC-Glauber model gives a correlation between eccentricity and multiplicity with positive slope ($\sim0.03 \pm 0.001$ ).

The opposite signs of the slopes in deformed U+U and Au+Au collisions in the two-component MC-Glauber model are a consequence of the $N_{\rm coll}$ dependence of the multiplicity and the opposite deformation of the U (prolate) and Au (oblate) nucleus. The anti-correlation between the 
ellipticity and the multiplicity in U+U collisions occurs because tip-tip configurations have large $N_{\rm coll}$ and small $\varepsilon_2$, while side-side configurations have small $N_{\rm coll}$ and large $\varepsilon_2$. In Au+Au collisions,  the oblate deformation leads to a correlation between $N_{\rm coll}$ and $\varepsilon_2$. We note that the qualitatively different MC-Glauber results in ultra-central collisions for the prolate U+U and oblate Au+Au geometries were also seen in earlier studies of these systems~\cite{Voloshin:2010ut}.

In the IP-Glasma model, there is no simple proportionality between the produced particle number and the number of binary collisions.
As noted in Section \ref{sec_def_ipg}, the multiplicity depends on $Q_s^2$, the transverse overlap area $S_\perp$ and the strong coupling constant $\alpha_s$. The relevant $Q_s$ for particle production is the smaller of the two from each nucleus at every position in the transverse overlap area. The (small $x$) coherence implicit in the model leads to a weaker ``thickness" dependence of the multiplicity unlike the two-component MC-Glauber model in which the thickness enters through $N_{\rm coll}$. This leads to a weaker anti-correlation between the multiplicity and the eccentricity in the IP-Glasma model relative to the two-component MC-Glauber model.

\section{Summary and conclusion}
\label{sec_sum}
We studied in this paper Cu+Au, U+U and Au+Au collisions at RHIC in the IP-Glasma framework. We presented results for single inclusive  multiplicities and multiplicity distributions as well as eccentricities $\varepsilon_n$, and the $\varepsilon_2$ event-by-event fluctuations. 

We do not see a large difference between the multiplicities per participant pair for central random and tip-tip configurations of U+U collisions in the IP-Glasma model, in contrast to the two-component MC-Glauber model. This indicates that the effect of a larger $Q_s$ in tip-tip configurations is largely compensated by a smaller overlap area.

The centrality and multiplicity dependence of the ellipticity $\varepsilon_2$ show significant sensitivity to the collision systems. 
In Cu+Au collisions $\varepsilon_2$ drops faster, because an average round shape is reached at a lower $dN_{\rm ch}/d\eta$ than in the heavier systems.
Due to the prolate deformation, ellipticities are generally smaller in tip-tip U+U collisions compared to random orientations, especially in the most central collisions.
The fluctuation-driven moments of eccentricities, particularly $\varepsilon_3$ and $\varepsilon_5$ are found to be very similar for different systems, while $\varepsilon_4$ shows a weak sensitivity to the colliding nuclei. 

We presented a comparison of the event-by-event distributions of ellipticities for different systems in the IP-Glasma model.
In the $20-30\%$ central bin studied, Cu+Au collisions were found to produce a wider distribution of $\varepsilon_2$ than the larger collision systems.
In this centrality bin, the IP-Glasma model yields a narrower distribution than the MC-Glauber model for Au+Au and U+U collisions. For very central collisions ($0-10\%$), this difference is negligible. 
As previously shown for event-by-event distributions of anisotropic flow coefficients at the LHC, similar measurements at RHIC energies can distinguish between models with different multi-particle production mechanisms. Our results suggest that extraction of event-by-event distributions for the collision systems studied here will help further in discriminating between models.

Finally, we presented results exploring the correlation between ellipticity and multiplicity for ultra-central events (events with a small number of spectator neutrons). IP-Glasma model computations show a much weaker anti-correlation between ellipticity and centrality than the two-component MC-Glauber model for U+U collisions. This striking difference follows from the different mechanisms of particle production in the two models. In particular, the explicit $N_{\rm coll}$ dependence of multiplicities in the two-component MC-Glauber model, which is absent in the IP-Glasma model, leads to very strong anti-correlations for prolate nuclei. A qualitative difference between the two models is seen for $0\!-\!1\%$ Au+Au collisions. A positive correlation between ellipticity and multiplicity is seen in the MC-Glauber model. In contrast, the IP-Glasma yields a weak yet distinct anti-correlation. 
 
Experimental measurements of correlations between ellipticities and multiplicities in ultra-central collisions, that are most significant for deformed nuclei, can therefore clearly distinguish between different models of multiparticle production. These will provide much needed constraints on the theoretical description of the initial state in heavy-ion collisions.

\vspace{20pt}

\section*{Acknowledgments}
We thank Subhasis Chattopadhyay, Roy Lacey, Hiroshi Masui, Paul Sorensen and Hui Wang for interesting discussions. This research used resources of the National Energy Research Scientific Computing Center, which is supported by the Office of Science of the U.S. Department of Energy under Contract No. DE-AC02-05CH11231, the DRONA and PRAFULLA cluster of Computer Division and the LHC grid computing centre at the Variable Energy Cyclotron Centre, supported by the Department of Atomic Energy, Government of India. BPS and RV are supported under DOE Contract No. DE-AC02-98CH10886.

\vspace{-0.5cm}
\bibliography{spires_pt}

\end{document}